# POSSIBLE ASTRONOMICAL REFERENCES IN THE PLANNING OF THE GREAT HOPEWELL ROAD

Giulio Magli
Dipartimento di Matematica del Politecnico di Milano
P.le Leonardo da Vinci 32, 20133 Milano, Italy.

*The possible existence of astronomical references in the planning of the so-called Great Hopewell Road, a 90 Kilometres straight road composed of two parallel earthen embankments which, according to recent surveys, likely connected the Hopewell ceremonial centres of Newark and Chillicothe, Ohio, are investigated. It turns out that a very peculiar, although simple, set of astronomical alignments took place in connection with the road during possible periods of its construction. The possibility of a non-fortuitous connection is thus proposed and analysed.*

## 1. Introduction

The culture usually referred to as Hopewell (100 BC –AD 400) flourished as a direct continuation of the previous cultures, with a smooth transition which can be located in the centuries between 500 BC to AD 200. Among the most characterizing features of Hopewell culture are the elaborated pottery, highly refined obsidian tools, copper and mica artefacts and, as is well known, the impressive, monumental earthworks which characterize their ceremonial centres (for an up to date survey on Hopewell culture see Charles and Buikstra 2006). Many such earthworks are lost today, but they were documented in the 19 century by the survey carried out by Ephraim Squier and Edwin Davis (1848). In addition, traces of the works can be found through aerial photography and other non-intrusive techniques.

It is known that many Hopewell earthworks and especially those of geometric form did not have practical functions, like defensive walls or boundaries of villages. They were instead connected with the religious world of the Hopewell; further, the absence of extended human settlements nearby as well as the analysis of the archaeological findings point to an interpretation of the concentrations of earthworks as pilgrimage centres (Lepper 1996).

In recent years, Archaeoastronomy has proved to be a powerful tool in order to better understand the symbolic contents of many such monuments. This holds true, in particular, for the enigmatic "circle+octagon" structures located at Newark and at High Bank respectively (Figs. 1 and 2). These earthworks are located in Ohio, separated by a distance of 90 kms (the High Bank complex, near Chillicothe, is barely visible today, while the Newark one is well preserved), and show striking similarities in design. Indeed, they are composed of a huge circle – around 320 meters of diameter, identical in dimensions in both sites - connected with an octagon. The High Bank octagon is smaller with respect to Newark's, but the disposition in plan is identical, the unique difference being in the orientation, because the main axis of the High Bank circle-octagon structure is oriented orthogonally with respect to the Newark one (we shall come back to this fact in section 3).

Both at High Bank and at Newark, alignments to the solar as well as to the lunar standstills have been documented (Hively and Horn 1982, 1984) and those of Newark have been further confirmed by recent research (Mickelson and Lepper 2007). In particular, the main axis of Newark (extending from the center of the circle and bisecting the octagon) and the longitudinal (i.e. that orthogonal to the main one) axis of High Bank are oriented to the maximal northern standstill of the Moon.

The Newark and the High Bank octagons are the unique Hopewell earthworks exhibiting such geometry. Their connected circles are identical, and their main axis are oriented in orthogonal directions. All these facts strongly put the case for the two structures to be strictly related, in spite of their considerable distance. In recent years, an important archaeological discovery has strengthened this possibility. Indeed looking to 19 century maps of the Newark earthworks, especially the one drawn by James and Charles Salisbury in 1862, it can be noted that an ancient road, composed of two parallel earthen walls, runs to the southwest. The Salisbury brothers traced this road for some kilometres, and proposed that it likely went all the way down to Chillicothe. During the course of the years, various attempts have been made to trace this road; finally, archaeologist Brad Lepper succeeded in recovering (either on the ground, or in aerial photographs) various parts of it (see Lepper 1995, 2006). In particular, a well documented segment is located 26 km south of Newark, and another one runs near the terminus at Chillicothe. Lepper was thus led to the conclusion that a straight road, composed of two parallel earthen walls at least one meter high and separated by 60 meters, oriented 31° 25' west of south if seen from Newark, once connected the two sites, a conclusion further strengthened by a recent LIDAR analysis (Romain and Burks 2008).

The road is very likely to be interpreted as a ceremonial pathway – perhaps a pilgrimage route – which Lepper calls the Great Hopewell Road. The name is certainly well chosen because the construction of such a straight road, as we shall see in next section, must have implied considerable technical difficulties. The aim of the present paper is to analyse the possible existence of non-casual astronomical references in the tracing of such a road.

## 2. Straight roads in antiquity and astronomy

Straight roads are the fastest way to connect two points on a *plane* surface. Of course, since the earth is round, no straight lines exist at all on it, however, provided that the two ends of a road are not too far apart, the line which can be traced starting from one end and pointing to the final destination does not differ, in practice, from the true shortest path between the same points (i.e. the arc of that great circle which passes through them).

Minimization of the distances was thus, certainly, a key reason in the project of the majority of the straight roads built, for instance, by the Romans and the Inca. However, this is not the end of the story, since the existence of a symbolic content in tracing straight paths is well documented in many cases, including for instance the Anasazi roads we shall mention later on, and the Maya sacred roads: one example is the causeway which connected Yaxuna and Coba'. The road, about 100 Kms length, runs along subsequent straight segments, the first one, starting at Yaxuna', is nearly 60 Kms long (Villa Rojas 1934); recently, archaeological investigation has argued for the existence of a even longer, peninsular-wide Maya road running east-west for approximately 300 Kms from the ancient city of Tihó (modern Mérida) to Puerto Morelos, on the east coast (Mathews 1999).

From the technical point of view, the problem of tracing a road composed by straight segments is readily solved. Indeed, it suffices to use fire signals during the night, or smoke or mirror signals during daylight hours, day by day, during the construction. However, this procedure works well only if the length of the segment is of the order of some kilometres. Indeed, again, the earth is round, and the visible horizon (``how far it is possible to see, due to the earth roundness'') is very near. It is an easy exercise of trigonometry to show that the distance in kilometres at which an object of ``zero`` height can be seen from an height of H meters approximately equals the square root of 13H kilometres. Thus, for a person 1.70 meters high, the visible horizon is about 5 km; from a tower 20 meters high it is of the order of 16 km. In the case in which the sight point is not at zero height, the two horizons sum up; thus the top of a tower 20 meters high can be seen at a distance of 21 kms by a person 1.70 meters high. Practically, this implies that, if the two ends of a straight road are at a distance of more than 10-20 kms, it becomes extremely difficult to fix the direction of the road by a single measure; of course, the use of subsequent aligned segments is also possible (each based, for instance, on three or more inter-visible fire signals) but the possibility of errors increases considerably. In spite of such difficulties, however, "perfectly" straight roads do exist in antiquity; further, even straight and very long roads connecting *previously chosen points* do exist. A wonderful example is to be found, as is well known, in the *Regina Viarum*, i.e. Via Appia. Constructed by the Roman consul Appio Claudio around the year 312 BC, it was originally built to connect Rome with Capua. The first part of the road deliberately led straight to the already existing town of *Tarracina*, today's Terracina, and was composed of two perfectly straight segments, one leading from Rome to the hill of Collepardo, at 36 Kms from Rome, which was certainly used as an intermediate post, and the other (skewed around two degrees with respect to the first) leading from Collepardo to Tarracina, as long as 62 kms.

It is extremely likely that astronomy was used in tracing the Via Appia (this argument is currently under active investigation by the author). The use of an astronomical method is, indeed, already very well documented in another case of Roman roads, that of Britain (Ferrar and Richardson 2003). For instance, the road called *Fosse Way* linked Exeter in South West England to Lincoln in the East Midlands, with a sequence of straight sections. Built in the second half of the first century AD, its length is 292 kilometres, and its errors of alignment are such that the whole road remains inside a strip of ten Kilometres in width. The azimuth of the road is close within 1% to the arc whose tangent is the rational fraction 3/5. The method which was used is, thus, very probably the following. First, a grid oriented to the cardinal points through astronomical observations was constructed point by point. Second, the road was traced as the hypotenuse of a right triangle in which the two short legs were integer numbers (in this case 3 and 5; a tendency to the use of Pythagorean triangles – i.e. right triangles with all the legs integers, as (3,4,5) – is documented as well in other examples). This simple but very clever method, if applied rigorously, is very precise and, further, it has a very important feature: the engineers did not need the use of dimensional units, such as measures of length, or angles. Indeed, given an arbitrary unit, the triangle which generates the orientation of the road with respect to the cardinal points can be reproduced in any desired scale (very likely, huge centuriations such as the famous Tunisian one of 26 AD were carried out in a similar way, see Decramer and Hilton 1998).

Astronomical orientation is well documented in another famous case of a straight road, namely the so called Great North Road . It is a long ceremonial route, having a width of 9 meters, constructed by the Anasazi around 1100 AD. The road starts near the north rim of Chaco Canyon and ends north-west at Kutz Canyon through three straight segments; in particular, the second one (that leading to the so-called Pierre's Complex) runs with less than ½ of a degree of error to true north for 16 kms (Sofaer et al 1989; for the interpretation of the Anasazi roads see Lekson 1999 and references therein). Precise methods for the astronomical determination of geographic north were thus known to the Anasazi, as it is made clear also by the orientation to the cardinal points of many buildings in Chaco Canyon, such as the famous Pueblo Bonito (Sofaer and Sinclair 1986), although it is difficult to ascertain if they used a solar method or a stellar one.

## 3. Possible astronomical references in the Great Hopewell Road

The problem of tracing a very long (tens of kilometres) straight road is thus extremely difficult if it has to be solved with high accuracy, and it is even more difficult if the ends are chosen *a priori.* In the case of the Hopewell road, we actually cannot be sure if the road was traced between two existing sites, or if one of the two was constructed deliberately at the end of the straight path. Curiously, at a first glance one is led to think that this is the most likely explanation, and that perhaps it was Newark the last to be constructed, because it is the northernmost of the great Hopewellian ceremonial centres. However, at least in the opinion of this writer, there is little doubt that the High Bank "circle+octagon" was constructed having in mind, at least, the project of Newark. In fact, the intentionality of the lunar alignments at Newark can hardly be disputed. Since the lunar maximum amplitude with respect to due east at this latitude is less than 45°, the main axis at High Bank is *not* oriented to the southern maximum standstill of the moon, but furter south (the main axis does not, actually, exhibit any recognisable astronomical alignment). However, there can be little doubt that also this structure was anyhow conceived with astronomical alignments to the solar and to the lunar standstills; in particular, the alignment symmetrical with respect to the east-west line to the Newark main axis is realized by the direction between the east corner of the octagon and the point were the main axis crosses the circle. What is important to notice here is that, as a consequence, a likely explanation of the fact that the High Bank octagon is smaller with respect to the Newark's one is, that the builders had to respect three "rules": 1) lunar alignments specular to Newark's 2) 90° rotation of the structure 3) identical geometry. It is easy to see, that the solution of this problem is to *rescale* the dimensions of the octagon consequently (see Hively and Horn 2006 for a statistical approach to this and related issues). Therefore, for reasons *not* known to us, we can be relatively sure that the 90° skew was a rule governing the construction of the site (there is, indeed, no recognizable feature of the territory which can justify the different orientation for practical reasons). Interestingly enough, an analogous rule seems to have governed the alignment of the Great Road. Indeed, inspecting the direction of the road (and thus the azimuth 31° 25' from High Bank to Newark, and 211° 25' from Newark to High Bank) it turns out that the direction *orthogonal* to the road aligns with impressive precision to the direction of the setting sun at the summer solstice (or to the opposite one, to the rising sun at winter solstice). In fact, assuming a flat horizon, in 100 AD the setting azimuth of the sun at Midsummer was 301° 30'. However, how can such an alignment - measurable for a few days during each year - be carried on with precision during the construction of the road? A possible solution is the use of the rising and/or setting positions of bright stars. Due to precession, such positions vary year by year (indeed, day by day), but of course the variation is very small. Given the azimuth of the road, it is therefore possible to inspect the period within which the road was constructed (for safety reasons, we consider here a broad period lasting form 250 BC to 200 AD, but see the discussion below) and check if there were stars rising and/or setting in alignment with the road. Actually, we have the following (see again Fig. 3):

- At azimuth 211° 25', an observer looking south-west with a flat horizon would have seen the bright star Fomalhaut, of the constellation Piscis Austrinus, setting in alignment with the road. The alignment is optimal around 250 BC. At this date, Fomalhaut was setting at an azimuth of 211° 25' with an altitude of 1° (since the magnitude of this star is around 1, the last altitude of visibility can be estimated to be 1°, see e.g. Aveni 2001).

- At azimuth 31° 25', an observer looking north-east would have seen the bright star Capella, of the constellation Auriga, rising in alignment with the road. The alignment is optimal at a date around AD 100.

**4. Discussion**

The above mentioned facts open the possibility that the road was constructed in such a way to be oriented 90° with respect to the summer solstice sunset, and that this alignment was traced along the tens of kilometres of the road by using bright stars. One possibility is the setting of Fomalhaut, and this would put the construction near the earlier possible date, in the transitional pre-Hopewell period; another possibility is the rising of Capella, bringing the date of construction around 100 AD. Actually, the date of construction of the road is not known with certainty, and it may be that the settlements were connected by a roadway much before its "monumentalization". However, the hypothesis of a plainly Hopewell work seems to be favoured from archaeological data (Lepper 1995); further, recent excavations by N. Greber at Chillicothe (cited in Lepper 2005, p.141) uncovered wooden remains which have been carbon-dated around 100 AD.

Of course, the author is well aware that the alignments proposed in this paper might be just due to a chance. There is no reasonable way to apply a statistical analysis to a single object such as the Great Road, but it may at least be recalled that another example of astronomically oriented Hopewell road is already documented, it is the so-called Marietta *Sacra Via* at the Marietta Works in Washington County, a 200 meters long "avenue", consisting of a set of parallel walls that flanked an excavated path, which points quite precisely to the winter solstice sunset (Romain 2000). Stellar alignments have been documented as well in Hopewell earthworks, in particular at Mounds State Park, Anderson, Indiana (Cochran & McCord 2001).

**Acknowledgements**

The author warmly thanks Bradley Lepper and Michael Mickelson for their useful comments and a careful reading of the first version of this paper .

Figure captions

Fig. 1
Plan of Newark earthworks, from Squier and Davis.

Fig. 2
Plan of High Bank earthworks, from Squier and Davis.

Fig. 3
Schematic orientation diagram for the rising of Capella and the setting of Fomalhaut at Newark between II century BC and I century AD. The Great Hopewell road is indicated by the thick line: Fomalhaut sets in alignment with this line in II century BC, then its setting point slowly moves westward with respect to the road, while the rising point of Capella approaches the road direction. Capella rises in alignment around 100 AD, then also this alignment starts to be lost due to precession, since Capella will rise further and further north with respect to the road. The summer solstice sunset- winter solstice sunrise line is also shown, skewed 90° with respect to the road.

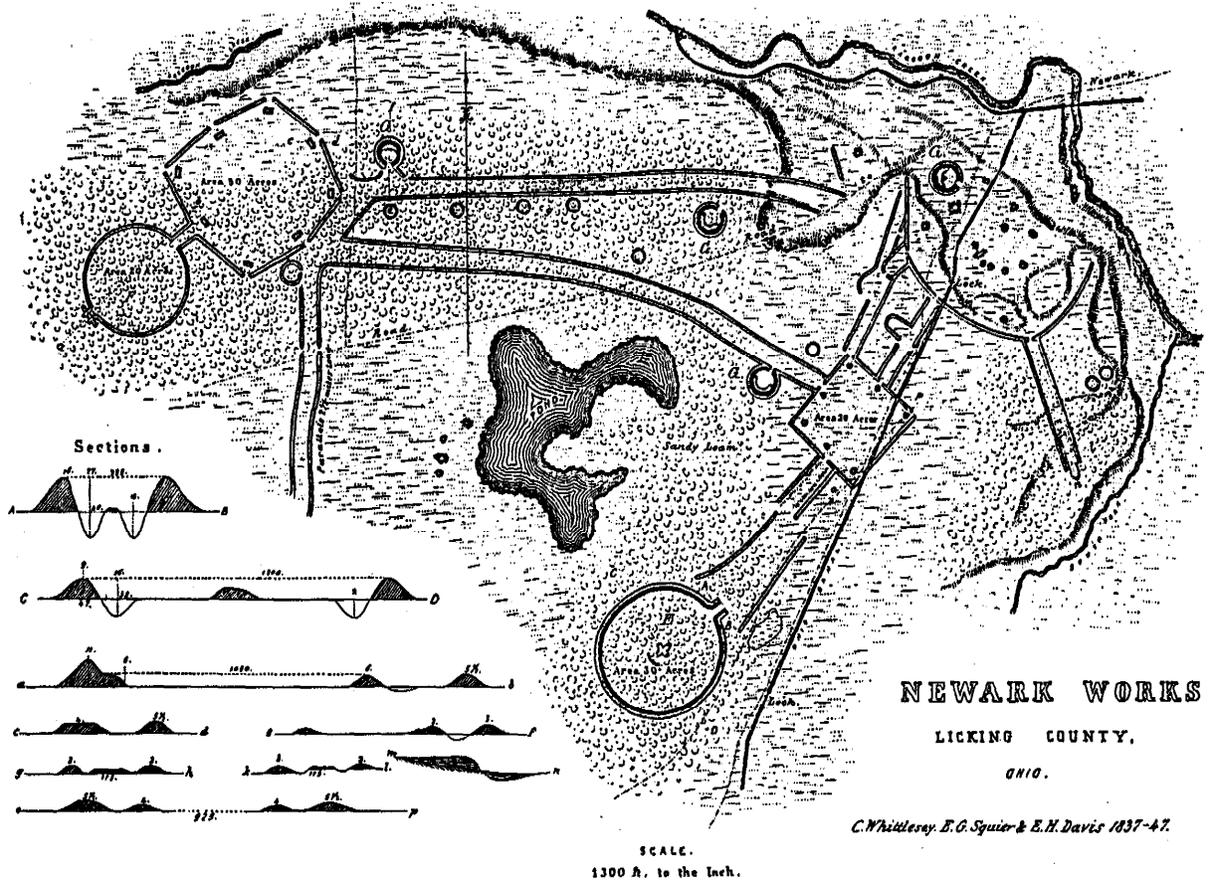

Fig. 1

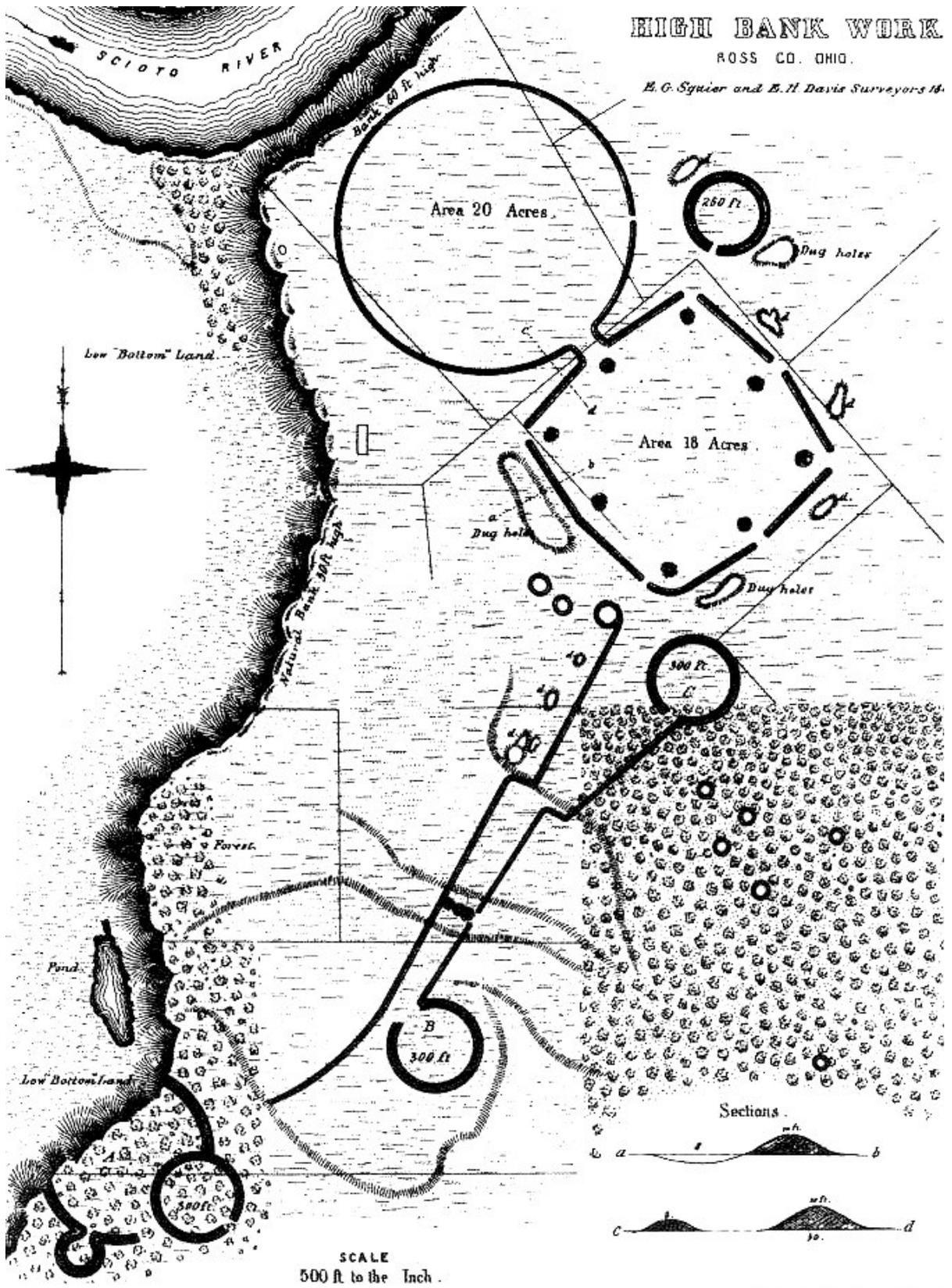

Fig. 2

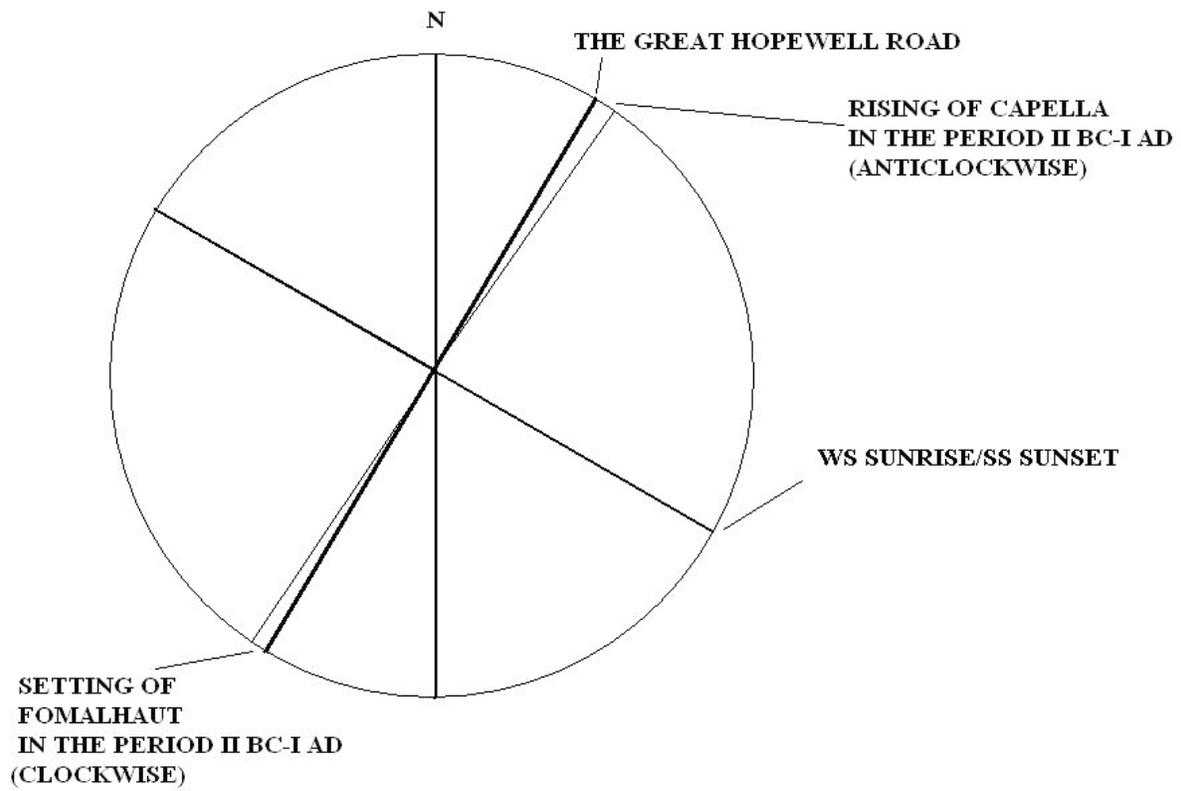

Fig. 3